\documentclass[twocolumn]{aa}
\usepackage{graphicx}
\usepackage{bm}
\usepackage{psfig}
\usepackage{epsfig}
\usepackage{txfonts}
\usepackage{natbib}
\bibpunct{(}{)}{;}{a}{}{,} 
\def\Msun{\hbox{$M_\odot$}}

\newcommand{\Zb}{\overline{Z}}
\newcommand{\mb}{\overline{m}}
\newcommand{\me}{m_{\rm e}}
\newcommand{\BHphi}{\hat{\bm{\varphi}}}
\newcommand{\BHtheta}{\hat{\bm{\theta}}}
\newcommand{\BHi}{\hat{\bm{\imath}}}
\newcommand{\BHj}{\hat{\bm{\jmath}}}
\newcommand{\BHk}{\hat{\bm{k}}}
\newcommand{\BHr}{\hat{\bm{r}}}
\newcommand{\BHu}{\hat{\bm{u}}}
\newcommand{\BHw}{\hat{\bm{w}}}
\newcommand{\gB}{g_{_{\rm B}}}
\newcommand{\half}{{\textstyle\frac{1}{2}}}

\newcommand{\cm}{{\mathrm{cm}}}
\begin{document}
   \title{Accretion in dipole magnetic fields:
          flow structure and X-ray emission of accreting white dwarfs}

   \subtitle{}

   \author{
        Jo\~{a}o Batista Garcia Canalle\inst{1,2},
        Curtis J. Saxton\inst{1,3},
        Kinwah Wu\inst{1},
        Mark Cropper\inst{1}
        \and 
        Gavin Ramsay\inst{1}
   }
   \authorrunning{Canalle et al.}
   \titlerunning{Accretion in dipole magnetic fields}

   \offprints{J.B.G. Canalle}

   \institute{
	Mullard Space Science Laboratory,
	University College London,
	Holmbury St.\ Mary, Dorking, Surrey,
	RH5 6NT, U.K.
\and
	State University of Rio de Janeiro,
	Rua S\~ao Francisco Xavier, 524/3023-D,
	CEP 20559-900, Rio de Janeiro, RJ, Brazil
\and
	Max-Planck-Institut f\"ur Radioastronomie,
	Auf dem H\"{u}gel 69, D-53121 Bonn, Germany
             }


   \abstract{
Field-channelled accretion flows occur in a variety of
  astrophysical objects, including T Tauri stars,
  magnetic cataclysmic variables and X-ray pulsars.
We consider a curvilinear coordinate system
  and derive a general hydrodynamic formulation
  for accretion onto stellar objects
  confined by a stellar dipole magnetic field.
The hydrodynamic equations are solved to determine
  the velocity, density and temperature profiles of the flow.
We use accreting magnetic white-dwarf stars
  as an illustrative example of astrophysical applications.
Our calculations show that
  the compressional heating due to the field geometry
  is as important as radiative cooling and gravity
  in determining the structure of the post-shock flow
  in accreting white-dwarf stars.
The generalisation of the formulation
  to accretion flows channelled by higher-order fields
  and the applications to other astrophysical systems are discussed.
   }

   \maketitle
%

\section{Introduction} 

Accretion is a common phenomenon in astrophysical systems 
  ranging from young stellar objects, interacting binaries, galaxies 
  to galaxy clusters. 
When the magnetic-field stress is larger than 
  the ram pressure of the accreting material, 
  the flow is confined to follow the magnetic-field lines.  
The accretion hydrodynamics in these systems 
  are therefore dependent on the magnetic-field geometry.  
 
Field-channelled accretion flow 
  can occur in young stellar objects 
  \citep[e.g.\ ][]{koenigl1991,hartman1994,li1996,koldoba2002,
  calvet1998,gullbring2000,lamzin1998,lamzin2001,romanova2003,stelzer2004},
  neutron star accretion from inter-stellar medium
  \citep[e.g.][]{toropina2003},
  and interacting binaries containing a white dwarf or a neutron star 
  \citep[e.g.\ ][]{elsner1977,ghosh1978,arons1993,lovelace1995,
  li1996b,kryukov2000,koldoba2002}.
In white-dwarf and neutron-star binaries 
  the magnetic field of the compact star can also be strong enough 
  to affect the cooling processes.     
Despite the fact 
  that these stars may have complex magnetic-field structures,  
  the dipole field component is important, 
  as it has a longer range than the higher-order field components 
  and dominates in the regions sufficiently far from the star.   
  
Here we investigate the accretion flow onto stellar objects 
  in the regime $B^2 \gg 8\pi \rho v^2$ 
  (where $B$ is the magnetic field, 
   and $\rho$ and $v$ are the density 
  and velocity of the accreting material respectively), 
  so that the flow is strictly confined by the magnetic field.    
We apply a curvilinear coordinate system 
  that is natural to the dipole-field geometry  
  for the hydrodynamic formation.  
We solve the hydrodynamic equations 
  for an accretion flow   
  in which the cooling function 
  has a power-law dependence on the temperature and density         
  and thereby obtain the flow-velocity, density and temperature profiles.  
  
We apply the model to  the post-shock flow
  in magnetic cataclysmic variables (mCVs),
  which are close binary systems containing a magnetic white dwarf 
  accreting from a red-dwarf companion star 
  \citep[see ][]{warner1995}.
We calculate the temperature, velocity and density structure 
  of the post-shock emission region
  and model the X-ray line and continuum emission.
We compare our results to those obtained by the plane-parallel model
  \citep{chevalier1982,wu1994a,wu1994b,cropper1999}
  which is generally used in spectral analysis of X-rays emitted from mCVs.

The paper is organised as follows: 
In \S2 we derive the hydrodynamic equations in curvilinear coordinates; 
  in \S3 we present the treatment of the boundary conditions 
  and discuss briefly the numerical scheme that we use; 
  and in \S4 to \S6 we show an illustrative example 
    --- accretion onto a magnetic white dwarf stars ---
   and present hydrodynamic structure and spectral calculations. 
In \S7 we consider the use of this formulation
  in some other astrophysical applications.
Our summary and conclusions comprise \S8.
  
\section{Formulation} 
    
\subsection{Coordinate systems} 

\begin{figure}
\begin{center}
\psfig{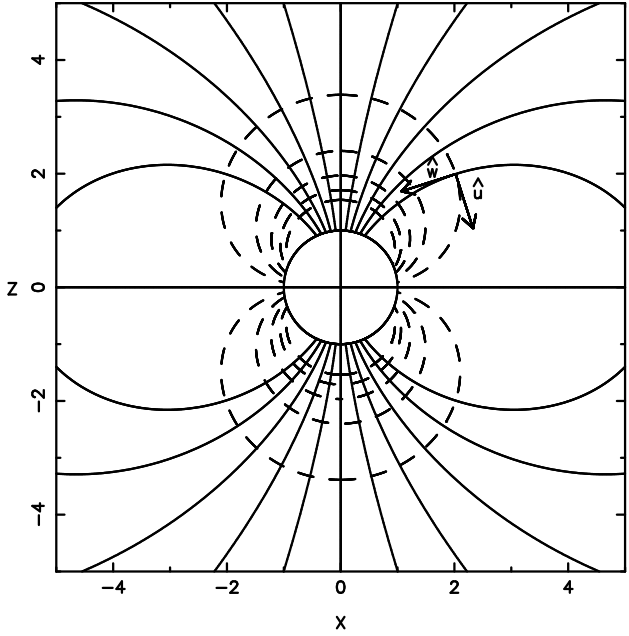} 
\end{center}
\caption{ 
The projection of the ({\it u, w, $\varphi$}) cordinates   
  on the {\it xz}-plane.   
The stellar magnetic dipole is located at the origin 
  and the symmetry axis of the dipole is the {\it z} axis.   
The central circle respresents the surface of the star.  
The solid lines are the constant-$u$ curves 
  and the dashed lines are the constant-$w$ curves.     
The unit vectors $\hat{\mathbf u}$ and $\hat{\mathbf w}$ are the unit normals 
  to the constant-$u$ and constant-$w$ curves respectively. 
The $\varphi$ coordinate, 
  whose normal is perpendicular to the {\it xz}-plane, 
  is not shown.  
 }
\label{dipolecoordinate}
\end{figure}

For the analytic study of field-channelled accretion flow,
  we choose a coordinate system   
  with one component along the magnetic-field lines.
In this representation,
  the flow is at most 2-dimensional (2D);    
  and it can often be reduced to 1-dimensional (1D) form
  when the system has a special geometric symmetry.
In the present study we consider accretion channelled by a dipole field,
  which is axi-symmetric and, it can be shown, essentially 1D.
We define a curvilinear coordinate system ({\it u, w, $\varphi$})
  for the dipolar field as follows.
The first coordinate $u$ is defined by the magnetic-field lines 
  generated by a point magnetic dipole.
The dipole is oriented in the $z$ direction 
  and located at the center of the accreting star 
  (the origin of the coordinate systems),  
  where $(x,y,x)$ are the usual Cartesian coordinates.   
As a static magnetic field is the gradient of a potential, 
  we use the equipotential curves of the dipole field
  as second coordinate $w$.
The third coordinate is the azimuthal coordinate, $\varphi$,  
  the same as that in the conventional spherical coordinate system 
  ({\it r, $\theta$, $\varphi$}).  
The unit vectors $\BHu$, $\BHw$ and $\BHphi$
  are orthogonal to each other, as shown in Fig.~\ref{dipolecoordinate}. 

In spherical coordinates, 
  the field lines and equipotential surfaces of a dipole field 
  satisfy $r = a_1 \sin^2 \theta$ 
  and $r^2 = a_2 \cos \theta$ respectively, 
  where 
  $a_1$ is the radius 
  at which the field line intercepts the mid-plane of the dipole
  and $a_2$ is the radius 
  at which the equipotential surface meets the polar axis.  
Thus, we have   
\begin{equation} 
  u = \frac{\sin^2 \theta}{r} \mbox{and}
\label{udipole} 
\end{equation}   
\begin{equation} 
  w = \frac{\cos \theta}{r^2} \ .
\label{wdipole} 
\end{equation}    
It follows that 
  $\theta = \cos^{-1} (wr^2)$
  and 
  the radial location $r$ is a function of $u$ and $v$,
  as given in (\ref{eq.radius.fn}) in the appendices.
Also, 
\begin{equation} 
  \left[{\begin{array}{ccc}
  x\ & y 
  \end{array}}\right]
  =
  \sqrt{u\,r^{3}}\,\,
  \left[{\begin{array}{ccc}
     \cos\,\varphi\ & \sin\,\varphi
  \end{array}}\right]\ ,
  \label{assign.xy}
\end{equation}
\begin{equation} 
  z = w\,r^{3} \ . 
  \label{assign.z}
\end{equation}     
(See Appendix~A for the coordinate transformations
and their derivations.) 

\subsection{Hydrodynamics equations}

The hydrodynamics of the flow is described by the conservation equations   
\begin{equation}
   \frac{\mbox{d}}{\mbox{d}t}\rho
   +\rho(\nabla\cdot\mbox{\bf v})=0 \ ,  
\label{continuity equation definition}
\end{equation}
\begin{equation}
   \frac{\mbox{d}\mbox{\bf v}}{\mbox{d}t}+
   \frac{1}{\rho}\nabla(P_{\rm e}+P_{\rm i})
   =\mbox{\bf g}+\mbox{\bf f}_{\rm rad} \ , 
\label{momentum equation definition}
\end{equation} 
\begin{equation}
   \frac{\mbox{d}P_{\rm i}}{\mbox{d}t}-\gamma\frac{P_{\rm i}}{\rho}
    \frac{\mbox{d}}{\mbox{d}t}\rho
   =-(\gamma - 1)\Gamma_{\rm ei} \ , 
\label{ion pressure equation definition}
\end{equation}
\begin{equation}
  \frac{\mbox{d}}{\mbox{d}t}P_{\rm e}
  -\gamma\frac{P_{\rm e}}{\rho}\frac{\mbox{d}}{\mbox{d}t}\rho
  =-(\gamma - 1)[\nabla\cdot(\mbox{\bf q}+\mbox{\bf F}_{\rm rad})
   -\Lambda_{\rm h}-\Gamma_{\rm ei}]  \ , 
\label{electron pressure equation definition}
\end{equation}  
   where 
$\mbox{d}/\mbox{d}t\equiv \partial/\partial t+ \mbox{\bf v}\cdot\nabla$, 
  and $P_{\rm e}$, $P_{\rm i}$, ${\bf v}$ and $\rho$ 
  are the electron pressure, ion pressure, velocity and density. 
$\gamma$ is the adiabatic index of the gas,  
  $\mbox{\bf g}$ is the gravitational acceleration,
  $\mbox{\bf f}_{\rm rad}$ is the acceleration due to the radiative force,
  $\mbox{\bf F}_{\rm rad}$ is the radiative flux, 
  $\Gamma_{\rm ei}$ is the electron-ion exchange rate, 
  and $\Lambda_{\rm h}$ is the heating function. 

For stationary flows, the time derivatives are zero. 
For one temperature flows,  
  the electron and the ion energy equations can be combined 
  by eliminating $\Gamma_{\rm ei}$, 
  yielding  
\begin{equation}
    (\mbox{\bf v}.\nabla)P -  \frac{\gamma P}{\rho}(\mbox{\bf v}.\nabla)\rho
      =-(\gamma -1) \,[\, \nabla\cdot(\mbox{\bf q}+\mbox{\bf F}_{\rm rad})
     - \Lambda_{h}]  \ , 
\end{equation} 
  where $P = P_{\rm e} + P_{\rm i}$ is the total pressure.  
We ignore the details of radiative transfer for simplicity   
  and assume that the total effect of heating and energy loss 
  can be described by a cooling function $\Lambda$  
  that depends only on the local hydrodynamical variables.  
We also ignore the radiative force and set ${\bf f}_{\rm rad} = 0$.   
Under these assumptions, we have a set of three conservation equations   
\begin{equation}
   \nabla \cdot (\rho \mbox{\bf v}) = 0 \  , 
\label{nablarhov}
\end{equation}
\begin{equation}
   (\mbox{\bf v}\cdot \nabla) \mbox{\bf v} + \frac{1}{\rho}\nabla P 
   = \mbox{\bf g}  \ , 
\label{nablapg}
\end{equation}
\begin{equation}
   (\mbox{\bf v}\cdot \nabla) P 
   - \frac{\gamma P}{\rho}(\mbox{\bf v}\cdot \nabla )\rho 
   = -(\gamma -1)\Lambda   \ . 
\label{gammaonelambda}
\end{equation}  
This set of hydrodynamic equations is closed 
  after we specify an equation of state for the gas.    

\subsection{Dipole-field channelled accretion flow} 
  
As we have assummed by default that the flow follows the field lines, 
  the velocity has only the component along the $\hat w$ direction, i.e.\  
\begin{equation}
  \mbox{\bf v}= v \, \hat{\mathbf w}  \ . 
\label{vfifi}
\end{equation}  
If we ignore the effect of stellar rotation, 
  the acceleration is due to only the gravitational force, 
  which is radial, i.e. ${\bf g}= -g_{*}\hat{\mathbf r}$, where 
  $g_*=GM_*/R_*^2$,
   $M_{*}$ and $R_{*}$ and the mass and radius of the accreting star, 
   and $G$ is the gravitational constant. 
For accretion flow along a dipole stellar magnetic field, 
  the gravitational acceleration has two components, i.e.\ 
\begin{equation}
   \mbox{\bf g}= g_{u}\, \hat{\mathbf u} + g_{w}\,\hat{\mathbf w}   \ , 
\label{gfifi} 
\end{equation}
  in which  
\begin{equation}
   g_{u} = -g_{*}\, \frac{Q_{11}}{r(u,w)^{2}}  \ , 
\label{gu} 
\end{equation} 
\begin{equation}
   g_{w} = -g_{*}\, \frac{Q_{12}}{r(u,w)^{2}}  \ ,  
\label{gw} 
\end{equation}   
  where $r(u,w)$ is given 
  by equation (\ref{eq.radius.fn}).
The terms $Q_{11}$ and $Q_{12}$ above 
  are the elements of the transformation matrix ${\mathsf Q}$ 
  (given in Appendix~A).    
 
The hydrodynamic equations decomposed into the orthogonal components 
  in the ({\it u}, {\it w}, $\varphi$)-coordinate system are  
\begin{equation}
   \frac{\partial}{\partial w}(h_{1}h_{3}\rho\,v)=0 \ , 
\label{rhovzero}
\end{equation} 
\begin{equation}
 \frac{v}{h_{2}}\frac{\partial v}{\partial w} + \frac{1}{h_{2} \rho }
  \frac{\partial P}{\partial w}  =g_{w} \ , 
\label{pgw} 
\end{equation}
\begin{equation}
   \frac{v}{h_{2}}\frac{\partial P}{\partial w}
   - \frac{\gamma}{h_{2}} \frac{P\, v}{\rho}\frac{\partial \rho}{\partial w}
   = - (\gamma -1)\Lambda  \ .    
\label{minusgammaonelambda}
\end{equation}  
In the equations above $h_1$, $h_2$ and $h_3$ 
  are the metric elements of the ({\it u}, {\it w}, $\varphi$)-coordinate system,  
  and their explicit expression are derived in Appendix~A.   
As we have assumed that the flow strictly follows the magnetic-field lines 
  (and omitted an explicit consideration of the plasmas and MHD effects,    
  e.g.\ the tension of the field lines and the magnetic stress), 
  we can ignore
  the momentum and force components in $u$ and $\varphi$ directions.
The three relevant hydrodynamic equations in this study 
  are therefore 
  equations (\ref{rhovzero}), (\ref{pgw}) and (\ref{minusgammaonelambda}), 
  which are respectively the mass-continuity, momentum-conservation 
  and energy-conservation equations along the magnetic-field lines.    
 
Direct integration of the mass-continuity equation (\ref{rhovzero}) yields
\begin{equation}
   h_{1}\,h_{3}\,\rho\,v = C \ , 
\label{rhovc}
\end{equation} 
  where $C$ is a constant 
  (related to the local accretion rate and the geometry of the field) 
  to be determined.  
It follows that 
\begin{equation}
   \frac{\partial}{\partial w} (h_{1} \, h_{3} \, \rho \, v^{2})
   =h_{1}\, h_{3} \, \rho \, v \frac{\partial v }{\partial w} \ , 
\label{delvdelw}
\end{equation}
   and the momentum-conservation equation becomes   
\begin{equation}
   \frac{\partial}{\partial w} (h_{1} \, h_{3} \,  \rho \, v^{2}) 
    + h_{1} \, h_{3}\, \frac{\partial P}{\partial w}  
    =h_{1}\, h_{2} \, h_{3}\, \rho \, g_{w}  \ . 
\label{h3rhogw}
\end{equation}  

To simplify the mathematics we consider a variable $\xi$, 
  as in \citet{cropper1999}, which is defined as
\begin{equation}
   \xi \equiv v + \frac{P}{\rho \, v} \   ,    
\label{definition of xi} 
\label{def.xi}
\end{equation}  
  and use it to replace $P$ 
  as a dependent variable in the hydrodynamic equations. 
We also consider a geometrical function 
\begin{equation}  
  {\mathcal H}(u,w) \equiv \frac{\partial}{\partial w}\ \ln~(h_1 h_3)  \ ,  
\end{equation}   
  which describes the variation of the cross-section area  
  of the flow-flux tube. 
(Numerical values of ${\mathcal H}$ for different field lines are shown in
Appendix~B).   
We may then rewrite the momentum-conservation equation as  
\begin{equation}
   \frac{\partial \xi}{\partial w}
   = {{g_w h_2}\over{v}} +{\mathcal H}(\xi-v)
   \ .
\label{dxidw}
\end{equation} 
Using the mass-continuity and momentum-conservation equations  
  in terms of the ${\mathcal H}$ function and the variable $\xi$, 
  we rewrite the energy-conservation equation as 
\begin{equation}
  \frac{\partial v}{\partial w}
  ={{-h_2}\over{\gamma(\xi-v)-v} }
   \left[{
	{{\gamma-1}\over{C}}h_1h_3\Lambda
	+{{\gamma{\mathcal H}}\over{h_2}}(\xi-v)v
	+g_w
   }\right].
\label{dvdw}
\label{vzetav}
\end{equation} 
We now have a set of coupled differential equations  
  (equations\ \ref{dvdw} and \ref{dxidw})
  that determine the accretion flow channelled 
  by a dipole magnetic field.   
When the explicit form of the cooling function $\Lambda$ is given 
  and the boundary conditions are specified,  
  equations (\ref{dxidw}) and (\ref{dvdw}) 
  can be solved along each field line (constant $u$ and $\varphi$)
  to obtain the velocity, density and temperature profiles. 
  
\subsection{Cooling function}

In this study, we simply consider an ideal-gas law 
\begin{equation}
   P = \frac{\rho\,k_{B}\, T}{ \mu\, m_{H}}  \ , 
\label{kbtmh} 
\end{equation}  
  for the equation of state of the gas 
  (where $T$ is the thermal temperature of the gas,  
  $\mu$ is the molecular mass of the gas, 
  $m_{H}$ is the Hydrogen atomic mass 
  and $k_{B}$ is the Boltzmann constant)  
  and ignore the microscopic plasma effects 
  due to the magnetic field.  
In a variety of astrophysical situations,   
  the cooling function of an accretion flow   
  can be approximated by 
\begin{equation}  
  \Lambda = A \rho^{\alpha} T^{\beta}  \ ,   
\label{coolf}
\end{equation}      
  where $\alpha$ and $\beta$ are the power-law indices 
  for the density and temperature.  
When $\alpha = 2$ and $\beta = 1/2$, 
  $\Lambda$ is the cooling function 
  for the optically thin free-free emission, 
  which is an important radiative-loss process 
  in accreting compact stars 
  and in some young stellar objects.  
From the mass-continuity equation
  (\ref{rhovc})
  and the equation of state for the gas
  (\ref{kbtmh}),
  we can express the cooling function 
  in terms of the hydrodynamic variables used in our formulation: 
\begin{equation}  
    \Lambda = A \,\left(\frac{C}{h_1\,h_3\, v}\right)^{\alpha} 
  \left[{v(\xi-v)}\right]^\beta  \ . 
\label{rootvzetav}
\label{Lambda.explicit}
\end{equation}    
Expressing the cooling function in terms of the hydrodynamic variables, 
  the energy conservation is     
\begin{eqnarray}
  {{dv}\over{dw}}&=&
    {{-h_{2}}\over{\gamma(\xi-v) - v}}
     \biggl[\ (\gamma-1) A \biggl( \frac{C}{h_1 h_3}\biggr)^{\alpha-1} 
     \left(\xi-v \right)^\beta v^{\beta-\alpha}     
   \nonumber  \\ 
   & & \hspace*{4cm}
   +{{\gamma{\mathcal H}}\over{h_2}}v(\xi-v) + g_w\biggr]  \ . 
\label{eq.dvdw.alphabeta}
\end{eqnarray}    

\section{Transonic accretion onto stellar objects}   

For non-relativistic, spherical (Bondi-Hoyle type) accretion onto stellar objects,  
  there are two classes of physical solutions 
  \citep[see e.g.][]{frank}.
In the first class,  
  the flow is subsonic from infinity to the surface of the accreting star. 
In the second class, the flow is initially subsonic 
  and becomes supersonic at certain radius from the star. 
As it requires the flow velocity to be zero at the stellar surface, 
  the flow will become subsonic again via a shock, 
  which converts the kinetic energy of the flow 
  to the thermal energy of accreting material. 

The situation is the same for flows channelled by a dipole magnetic field,   
  although the dipole field lines, except the two originating from the poles, 
  are all closed at finite distances.   
\citep[See e.g.][for more details.]{koldoba2002}
In this study we consider the second class 
  and focus on determining the density and temperature structures  
  of the region between the shock and the star. 
We will show how to construct the shock boundary conditions 
  within the hydrodynamic framework that we use 
  and solve the hydrodynamic equations with these boundary conditions.   

\subsection{Boundary conditions}   
\label{section.bc}
 
We apply the Rankine-Hugoniot condition 
  to derive the shock boundary condition,  
  assuming that the pre-shock flow is supersonic and cold, 
  such that the Mach number ${\cal M} \rightarrow \infty$, 
  we have 
\begin{equation}
  V_{\rm s}= \frac{1}{4}V_{\rm pre} \ , 
\label{mach}
\end{equation} 
  (for an adiabatic index $\gamma = 5/3$),    
  where $V_{\rm pre}$ and $V_{\rm s}$ are the pre-shock 
  and post-shock velocities respectively. 
As a reasonable approximation,
  we may set $V_{\rm pre} = V_{\rm ff_{s}}$, 
  where $V_{\rm ff_{s}}$  
  is the free-fall velocity of the gas 
  before encountering the shock. 
We denote the free-fall velocity at the stellar surface as
\begin{equation}
   V_*=\sqrt{\frac{2 G\,M_{*}}{R_{*}}} \ ,  
\label{vffwd definition}
\end{equation}   
  and at any general position $(u,w)$
$V_{\rm ff}=V_*/\sqrt{r}$.
  Hence, we have the boundary condition
\begin{equation}
  V_{\rm s}
     = {{V_*}\over{4\sqrt{r_{\rm s}}}}  \ . 
\label{freefall}
\end{equation}   
(Here we use the conventions: 
  the shock is located at $r = r_{\rm s}$
  and the stellar surface is located at $r=1$.)   

If the collisional energy exchange between electrons and ions
is efficient compared to radiative cooling
then the electrons and ions have approximately equal temperatures.
The shock temperature is given by   
\begin{equation}
  T_{\rm s} =
   \frac{3}{8}\frac{\mu\,G\,M_{*}\,m_{H}}{k_{B}(R_{*}+x_{\rm s})} \   
\label{kts}\end{equation} 
  (see \citealt{frank})
  where the shock height above the stellar surface  
  $x_{\rm s} \equiv (r_{\rm s}-1)R_{*}$.
Recall the definition of the variable $\xi$, 
  equation (\ref{def.xi}),
  and we have  
\begin{equation}
  \xi_{\rm s}
  =V_{\rm s}+\frac{k_{B}T_{\rm s}}{\mu m_{H}V_{\rm s}} \ , 
\label{kbts}
\end{equation} 
  which implies that,
  regardless of the shock location $r_{\rm s}$,  
\begin{equation} 
 \xi_{\rm s}= V_{\rm {ff}_{\rm s}} \ .
\end{equation} 

At the stellar surface,  
  we have the stationary-wall boundary condition for the velocity, 
  that is $v=0$.  
The variable $\xi$ specifies the specific momentum flux, 
  and determining its boundary value at the stellar surface is less trivial, 
  as the flow is not necessarily perpendicular to the stellar surface. 
Nevertheless, knowing only the value of $v$ at the stellar surface
  together with the values of $v$ and $\xi$ at the shock is sufficient 
  to solve the hydrodynamic equations. 

\subsection{Modified hydrodynamics equations 
  and numerical technique}   

The stellar surface corresponds to $r=1$ in the spherical coordinates, 
  but the functional form of the stellar surface 
  in the $(u, w, \varphi)$ coordinates is not trivial.  
At the surface the independent variable $w$ 
  has different values for different magnetic field lines.  
The boundary value of $w$ for a field line  
  is determined (using equation (\ref{wdipole}))
  only after we specify the colatitude on the stellar surface 
  at which the field line is anchored
  (which we shall denote as $\theta_0$). 

As the boundary values of $w$ both at the shock and at the stellar surface  
  are not easily specified in terms of simple functions of $w$ and $u$,  
  we search for an alternative independent variable.   
The criterion for the variable is that 
  it increases or decreases montonically 
  in the region of interest. 
As shown above (\S\ref{section.bc})
  the velocity $v$ is better defined 
  at the shock boundary and the stellar-surface boundary. 
It should be smooth in the post-shock region  
  and is monotonic in at least in the region just beneath the shock 
  and in the region just above the stellar surface.     

Without losing qualitative generality, 
  hereafter we will use $\alpha = 2$ and $\beta = 1/2$
  (equation (\ref{Lambda.explicit}), describing Bremsstrahlung cooling)
  in our calculations.
Then the energy-conservation equation (\ref{eq.dvdw.alphabeta})
  can be expressed with $v$ the independent variable
  and $w$ the dependent variable,
\begin{eqnarray}
  \frac{d w}{d v} & = & - \frac{1}{h_2}
  \big[ \gamma (\xi - v) - v \big] \,v^{3/2}  
  \biggl[ \frac{(\gamma-1)A\,C}{h_{1}h_{3}}\sqrt{\xi-v}  \nonumber \\ 
 & & \hspace*{1cm} +\frac{\gamma {\mathcal H}}{h_{2}} (\xi-v)\,v^{5/2}  
  +g_{w} v^{3/2}  \biggr]^{-1}
  \ .
\label{gammazetavv} 
\label{dwdv}
\label{vzetav2}
\end{eqnarray}  

Applying the chain rule for differentiation, 
  we can also derive (from equations \ref{dxidw}) 
  another equation for the variable $\xi$, that is     
\begin{eqnarray}
  \frac{d \xi}{d v} & = & \left[ h_{2} \frac{\partial w}{\partial v}\right] 
   \left[ \frac{g_{w}}{v} + \frac{{\mathcal H}}{h_{2}}(\xi -v)  \right] \nonumber \\   
    & = & -\big[ \gamma(\xi-v)-v \big]   
   \left[ g_{w} v^{1/2}   + \frac{{\mathcal H}}{h_{2}} (\xi-v)v^{3/2} \right]  \nonumber \\  
  & & \hspace*{0.2cm} 
  \times \biggl[ \frac{(\gamma-1)A\,C}{h_{1}h_{3}} \sqrt{\xi-v}
   +\ \frac{\gamma {\mathcal H}}{h_{2}} (\xi-v)\,v^{5/2} 
   + g_{w}\,v^{3/2} \biggr]^{-1}  \ . 
\label{vcincomeio}   
\label{dxidv}
\end{eqnarray}
  
The shock-boundary value of $v$ is still not constant among the field lines
(i.e. for different choices of $u$). 
Consider a new variable 
   $\tau \equiv{v}/{V_{\rm f\!\!f}}$,
  where $V_{\rm f\!\!f}(r)$ is the local free-fall velocity 
  at the radial distance $r$ from the centre of the star, 
For the shock-jump conditions that we assume 
  (equations\ \ref{mach} and \ref{freefall}), 
  the variable $\tau$ has a fixed value at the shock  
  and is constant at 1/4.    
Thus, $\tau=0$ at the stellar surface 
  if $v=0$. 
The free-fall velocities  $V_{\rm f\!\!f}=V_*/\sqrt{r}$,
   and hence    
\begin{equation}
  v=\frac{\tau\, V_*}{\sqrt{r(u,w)}} \ ,  
\label{tau}\end{equation} 
  where $r(u,w)$ is given in equation (\ref{ruv}).   

The corresponding hydrodynamic equations,
 for $dw/d\tau$ and $d\xi/d\tau$,
 can be obtained 
 using the chain rule of differentiation with
\begin{equation} 
  \frac{d\tau}{dv} = \tau \biggl[ 
  {\frac{\sqrt{r(u,w)}}{\tau V_*}} 
    + \frac{1}{2}\biggl(\frac{r'_w(u,w)}{r(u,w)}\biggr) 
     {\frac{dw} {dv}}  \biggr]  \ , 
\end{equation} 
  and ${r'_w(u,w)}$ given in Appendix~A. 
 
The use of $\tau$ as an independent variable
  provides a convenient way to treat the boundary conditions
  at both the shock and the stellar surface.
However $\tau$ (like $v$) is not guaranteed to be monotonic in $w$
  for all systems and all choices of $u$.
Thus we cannot use it as the independent variable
  throughout the entire post-shock region.
In contrast, the coordinate $w$ is always monotonic along a field line,
  but it is not practically usable at the stellar surface,
  where the velocity gradient $\partial v/\partial w$
  approaches infinity.

Given that it is straightforward to use $w$ as the independent variable  
  in the numerical integration in the entire flow 
  except at the stellar surface, 
  we consider a hybrid numerical scheme in our calculations. 
Therefore we use $\tau$ or $v$ as the variable of integration
  in the vinicity of the stellar surface,
  $0\leq v\leq \delta v$ with $\delta \la 10^{-3}$,
  but we switch to integration in terms of $w$
  in the upper parts of the post-shock structure.
This algorithm allows us to solve the flow profile 
  along each field line.   


We calculate the pressure and velocity profiles of the post-shock structure
by numerically integrating a set of differential equations
between the upper and lower boundaries.
In regions where we treat $w$ as the independent variable,
we use equations (\ref{dxidw}) and (\ref{dvdw}).
Where $v$ or $\tau$ serves as the independent variable then
we use equations such as (\ref{dwdv}) and (\ref{dxidv})
or the equivalent with derivatives in $\tau$.

Neither boundary condition is completely specified from first principles:
the position of the shock (expressed in terms of
$w_{\rm s}$ or $r_{\rm s}$)
is not initially known.
Nor do we have foreknowledge of
the momentum flux at the stellar surface ($\xi_*$).
In practice we choose trial values of $(\xi_*,w_{\rm s})$
and test how the resulting profile conforms to the conditions required
at the opposite boundary.
In one method, we choose trial values of $\xi_*$,
and integrate upwards from the stellar surface until a shock is found.
We adjust $\xi_*$ until we match the conditions
$\xi_{\rm s} = V_*/\sqrt{r_{\rm s}}$
and $v_{\rm s} = \frac14 V_*/\sqrt{r_{\rm s}}$
simultaneously.
Alternatively, we may try values of $r_{\rm s}$ 
and integrate downwards from the shock to the point where $v=0$.
If $r\neq 1$ when $v=0$ then we adjust the trial value of $r_{\rm s}$
and retry.
These tests are applied iteratively in a root-finding routine.

\section{Accretion onto a magnetic white dwarf}

We now apply our formulation to an astrophysical system 
  as an illustration. 
We consider a simple case:  
  accretion onto a magnetised white dwarf, 
  in which free-free emission is the dominant cooling mechanism.  
This kind of system can be found 
  in intermediate polars and polars, 
  which are cataclysmic variables 
  containing a magnetic white dwarf 
  accreting material from a low-mass companion star 
  (for reviews of magnetic cataclysmic variables, 
  see e.g.\ \citealt{cropper1990}
  and \citealt{warner1995}).
The post-shock accretion flow 
  in magnetic cataclysmic variables 
  have been investigated by many workers   
  and analytic and semi-analytic results were obtained 
  (e.g. \citealt{aizu1973,chevalier1982,wu1994a,wu1994b, imamura1996,cropper1999,saxton1999,saxton2001}).
Their studies address various issues 
  such as the cooling processes and the effect due to gravity, 
  and the calculations either planar or spherical geometries were considered.  
(For reviews of the hydrodynamics of post-shock accretion 
  in magnetic cataclysmic variables, see \citealt{wu2000,beuermann2004}.)
Because of the assumed geometry, 
  these studies have not quantified 
  the effects of the curvature of field lines 
  in determining the hydrodynamic structure of shock-heated region.  
  
For the application of our formulation in the accretion onto white dwarfs, 
  we simply set $M_{*} = M_{\rm wd}$ and $R_{*} = R_{\rm wd}$, 
  where $M_{\rm wd}$ and $R_{\rm wd}$ and white-dwarf mass and radius, 
  in the hydrodynamic equations 
  and scale the variables accordingly.   
The white-dwarf mass and radius are not independent  
  --- when the mass is specified  
  one can calculate the radius by means of a mass-radius relation. 
In our calculations,  
  we adopt the \citet{nauenberg1972} mass-radius relation 
\begin{equation}
  \frac{R_{\rm wd}}{R_{\odot}}=
  \frac{0.0225}{\mu_{\rm wd}}\frac{\sqrt{1-(M_{\rm wd}/M_{3})^{4/3}}}
   {(M_{\rm wd}/M_{3})^{1/3}}  \ , 
\label{nauenberg equation}
\end{equation}
  where ${M_{3}} = {5.816}~{M_{\odot}}/{\mu_{\rm wd}^{2}}$
  is the Chandrasekhar mass limit.
We set the electron mean molecular weight parameter
  $\mu_{\rm wd} = 2.00$.
This is a good approximation for white dwarfs
  with He, C/O or Ne/Mg composition.

We consider a strong shock for the upper boundary condition 
  and a cool stationary wall for the lower boundary condition.  
The accreting material is an ideal gas with $\gamma=5/3$
  and approximately solar abundances \citep{anders1989}.
The mean ionic mass and charge are
  $\mb=2365.8 \me$
  and
  $\Zb=1.0999$ respectively,
  and $n_{\rm e}/n_{\rm H}=1.209$,
  $\overline{Z^2}=1.3912$,
  $\overline{Z^2/m}=6.007\times10^{23}\ {\rm g}^{-1}$.
We omit other radiative transport and microscopic effects 
  at the shock and the stellar-surface boundary 
  \citep[see e.g.][]{imamura1996,saxton2001,wu2001}.

\subsection{Bremsstrahlung-dominated flows}

If the white dwarf has a weak magnetic field
  or sufficiently high accretion rate
  then the cooling of the post-shock flow is dominated by free-free emission,
  and Cyclotron cooling is unimportant.
The cooling function is then 
\begin{equation}
   \Lambda = \Lambda_{\rm br} = A \rho^{2} \sqrt{\frac{P}{\rho}} 
\label{Lambda.ff}
\label{rhoprho}
\end{equation}
  with $A$ being a constant that depends on the composition of the plasma
  \citep[see][]{rybicki}.
For a purely Hydrogen plasma,
  $A=3.97 \times 10^{16} \gB$ in c.g.s.\ units
  where $\gB \sim 1$ is the Gaunt factor.
When we adopt $\gB=1.25$ and approximately solar abundances
  we have $A=5.06 \times 10^{16}$.

\subsection{Flows with Bremsstrahlung and Cyclotron cooling}

A natural extension of the case for accretion onto weakly magnetic white dwarfs 
  is the accretion onto strong-field systems
\citep[i.e.\ polars, see][]{cropper1990}
  the flow is cooled 
  by emitting both optical Cyclotron radiation and free-free X-rays.   
We can use a composite cooling function to describe the cooling process, i.e.\  
\begin{equation}  
  \Lambda = A \rho^2 \sqrt{\frac{P}{\rho}}\ 
   \left[{\ 1+\epsilon_s f(P,\rho,u,w) \ }\right]  
\end{equation}   
\begin{equation}
 f\equiv{{\Lambda_{\rm cy}}\over{\Lambda_{\rm br}}}=
 \left({\frac{P}{P_{\rm s}}}\right)^2 
 \left({\frac{\rho_{\rm s}}{\rho}}\right)^{\frac{77}{20}}
 \left({ {{h_{1\rm s}h_{3\rm s}}\over{h_1 h_3}} }\right)^{\frac{17}{40}}
 \left({ {B}\over{B_{\rm s}} }\right)^{\frac{57}{20}}
\label{def.f}
\end{equation}
  \citep[see][ and appendix~\ref{app.epsilons}]{wu1994b},
  where $\epsilon_{\rm s}$ is the parameter that determines the 
  relative efficiency of Cyclotron cooling, 
  and the subscript ${\rm s}$ denotes quantities evaluated at the shock. 
The penultimate factor in (\ref{def.f})
  appears because the Cyclotron cooling power
  depends on the cross-sectional area of the accretion stream,
  which is proportional to $h_1 h_3$.
The last factor describes the local magnetic field strength,
  given by
\begin{equation}
B(u,w) = { \sqrt{4-3ur}\over{2r^3} }
B_{\rm p}
\ ,
\end{equation}
where $B_{\rm p}$ is its value at the pole.

Then, we replace $\rho$ by $C/h_1 h_3 v$ and $P/\rho$ by $(\xi-v)v$ 
  throughout the $f$ cooling function,
  and substitute $\Lambda$ into equation (\ref{vzetav}).   
The resulting hydrodynamic equations can be solved 
  using the numerical scheme described in \S3.

\subsection{Simplification to previously published models}

We note that the formulation we derive 
  can reduce to the formulations obtained in the previous studies 
  under certain approximations and restricted conditions.  
Take, for instance, equations~(\ref{gammazetavv}) and (\ref{vcincomeio}).  
If we set $h_1 = h_2 = h_3 =1$, ${\mathcal H}=0$ and $g_{w} =0$, 
  and fix $\xi$ to be a constant 
  equal to the free-fall velocity at the white-dwarf surface,  
  then equation (\ref{vcincomeio}) vanishes 
  and equation (\ref{gammazetavv}) becomes the same as 
  that in \citet{chevalier1982} and \citet{wu1994a} for the planar flows. 
If we set $u=0$,
  $w=1/r^2$,
  $h_2=-dr/dw$,
  $h_1 h_3 = {\frac12}$,
  and ${\mathcal H}=0$
  then we obtain the same set of two equations of \citet{cropper1999}.

Comparing the formulation of \citet{cropper1999}
  with \citet{chevalier1982}
  and Wu (1994) 
  reveals that the former needs two equations to describe the flow
  while the later requires only one.
It is because in the presence of gravity 
  the quantity $\xi$, which describes the specific momentum flux,  
  is no longer conserved along the field lines.    
In our formulation, despite the presence of a dipole field, 
  two differential equations are sufficient 
  to determine the hydrodynamics,  
  as in the case of \citet{cropper1999}.
We still have two differential equations 
  when we set $g_{w}=0$ 
  in both equations (\ref{gammazetavv}) and (\ref{vcincomeio}).  

To further illustrate the importance of the effects 
  due to the field geometry, 
  in the next section
  we present the numerical results of calculations, 
  in which typical parameters 
  of magnetic cataclysmic variables are used. 
We will show that the results   
  can differ substantially 
  from those obtained by formulations  
  \citep[e.g.][]{cropper1999}
  without taking account of the geometric effects.

\section{Hydrodynamic structure}
\label{s.structure}

\begin{figure}
\begin{center} 
\includegraphics[width=8.4cm]{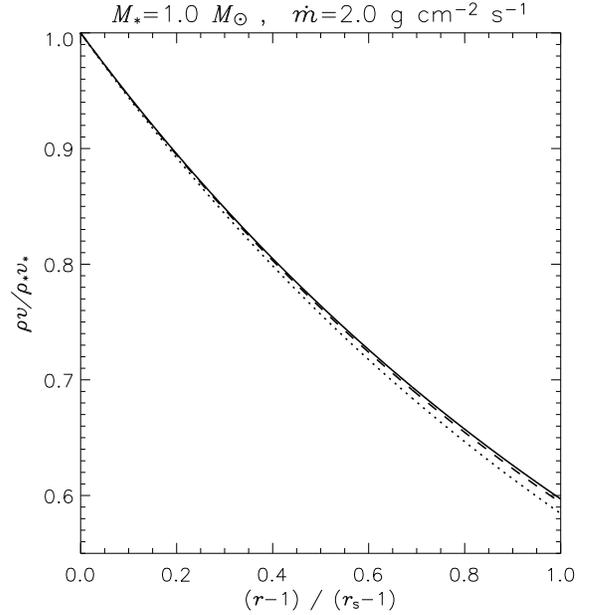}
\end{center}
\caption{
Variation of the local mass accretion rate $\rho v$ 
within the post-shock structure.
Different cases of the accretion colatitude,
$\theta_0 = 0^\circ, 18^\circ, 36^\circ$,
are marked by full, dashed and dotted lines respectively.
At the white dwarf surface we have
$\rho_* v_* \equiv \dot{m}$ by definition.
In this illustrative case
$\dot{m}=2.0\ {\rm g}\,\cm^{-2}\,{\rm s}^{-1}$,
$B_*=0$.
and $M_*=1.0M_\odot$.
The horizontal axis is 
the altitude above the white dwarf surface,
relative to that of the shock, $r_{\rm s}-1$.
}
\label{fig.rhov}
\end{figure}

The constant $C$ introduced in (\ref{rhovc})
  is obtained defining the value of $\rho v$
  at the bottom of the accretion column, $\rho_* v_*$.
Following \citet{cropper1999}
  we identify the accretion rate $\dot{m}=\rho_* v_*$,
  but in the present work the product $\rho v$ is not constant
  in the accretion column,
  since the area of the funnel, $\propto h_1h_3$,
  varies throughout the accretion stream.
In our standard illustrative cases, we use
  $\dot{m}=2.0\ {\rm g}\,\cm^{-2}\,{\rm s}^{-1}$,
  a white dwarf mass of $1.0M_\odot$,
  escape velocity
  $V_*=6.9\times10^{8}\,\cm\,{\rm s}^{-1}$
  and the accretion area is taken to be
  $3.9\times10^{15}\,\cm^2$,
  exactly $10^{-3}$ of the global surface area.
Figure~\ref{fig.rhov}
  shows how the product $\rho v$
  decreases with height within the column
  for several values of accretion colatitude, $\theta_0$.

\begin{figure}
\begin{center} 
\psfig{file=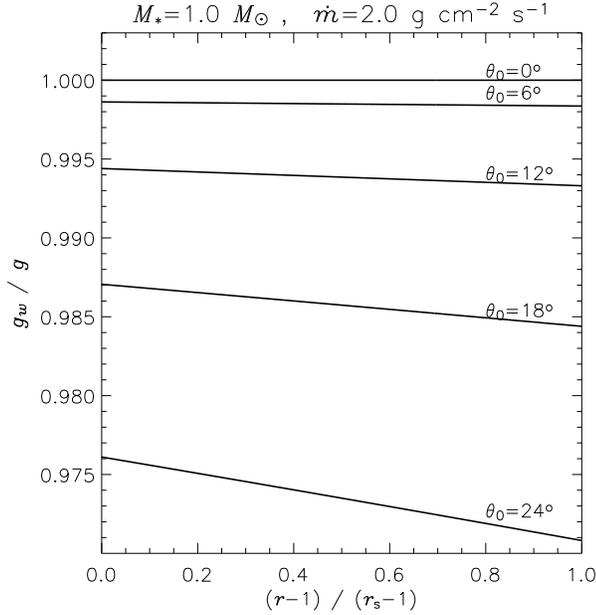,width=8.4cm}
\end{center}
\caption{
Variation of the component of the gravitational acceleration
tangential to the magnetic field line.
The five curves correspond to cases of
hot-spot colatitude $\theta_0=0^\circ, 6^\circ, 12^\circ, 18^\circ, 24^\circ$.
The parameters $(M_*,B_*,\dot{m})$ 
take the values used in Fig.~\ref{fig.rhov}.
}
\label{fig.gw.over.g}
\end{figure}

The gravitational acceleration component
  in the direction tangential to the magnetic field lines,
  which we call $g_{w}$ in (\ref{gw}),
  is a fraction of the total gravitational acceleration $g$.
Figure~\ref{fig.gw.over.g}
  shows the relation $g_{w}/g$.
Except for the special case of accretion onto the pole,
  $\theta_0=0^\circ$,
  the magnetic field lines are nowhere vertical to the stellar surface.
Thus the component $g_w$
  is not equal to the total acceleration on the surface of the white dwarf.

Scaled relative to the shock height and parameters of the accretion,
  there are several qualitative differences between
  the post-shock flows in the present dipolar model
  and the cylindrical accretion model of \citet{cropper1999}.

\begin{figure}
\begin{center} 
\psfig{file=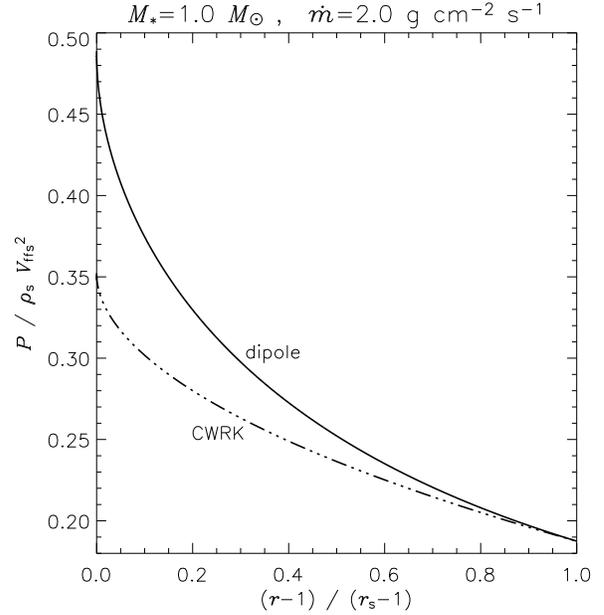,width=8.4cm}
\end{center}
\caption{
Post-shock pressure profiles
with the parameters ($M_*,B_*,\dot{m})$ used in
Fig.~\ref{fig.rhov}.
The vertical axis is scaled in terms of 
the free-fall velocity
($V_{\rm ff_s}$)
and density
($\rho_{\rm s}$).
immediately downstream of the shock.
The lower curve (CWRK) is the cylindrical accretion model.
The upper curve is the dipole model with $\theta_0=0^\circ$.
Varying $\theta_0$ has little effect on the appearance of this curve.
}
\label{fig.P.normal}
\end{figure}

The distribution of pressure is shown in
  Fig.~\ref{fig.P.normal}.
The dipolar accretion model results in
  proportionally higher pressures throughout the post-shock region.

\begin{figure}
\begin{center} 
\psfig{file=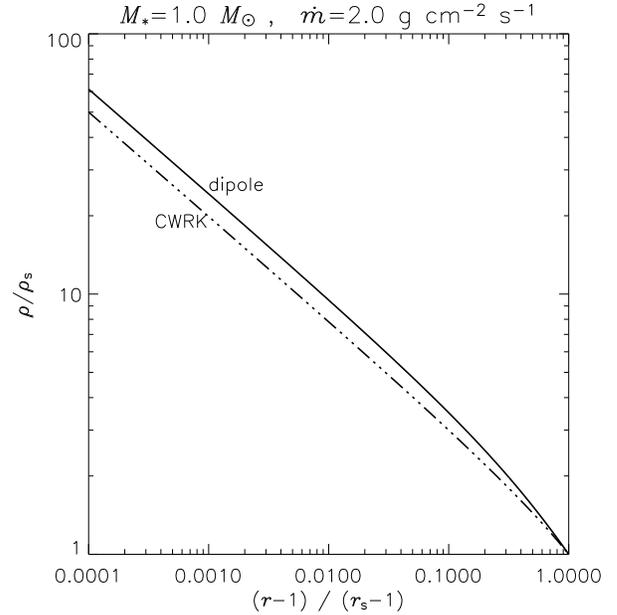,width=8.4cm}
\end{center}
\caption{
Density profiles of the accretion columns shown in 
Fig.~\ref{fig.P.normal},
scaled to the gas density immediately downstream of the shock
($\rho_{\rm s}$).
In the upper curve $\theta_0=0^\circ$,
and variation of $\theta_0$ has little effect.
The cylindrical accretion model (CWRK) is the lower
(dot-dot-dot-dash) curve.
}
\label{fig.log.rho}
\end{figure}

Corresponding profiles of the density structures are shown in
  Fig.~\ref{fig.log.rho}.
The dipolar funnel results in greater densities 
  throughout most of the post-shock region
  (relative to the gas density at the shock),
  than in the cylindrical accretion case.
As in previous models with a power-law expression
  for Bremsstrahlung cooling
  \citep[e.g. ][]{chevalier1982,wu1994a,cropper1999},
  the density necessarily increases asymptotically near the stellar surface.

\begin{figure}
\begin{center} 
\psfig{file=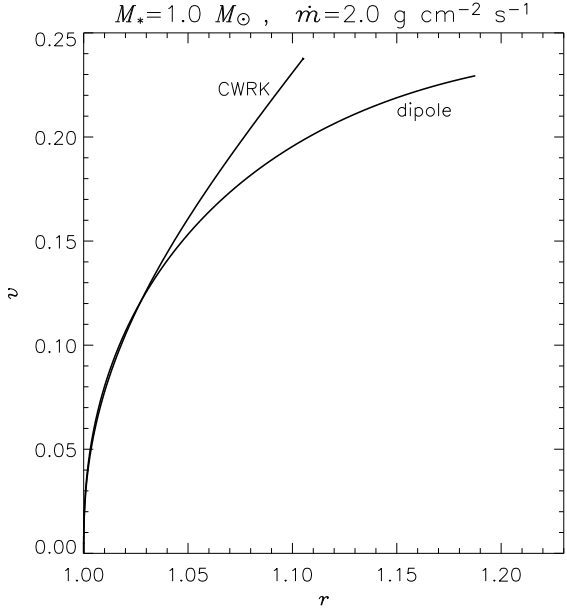,width=8.4cm}
\end{center}
\caption{
Velocity profile of the post-shock structure,
normalised to the escape velocity and the stellar surface, $V_*$.
The upper curve represents the model of Cropper et al. (1999).
The lower curve shows the effects
of accretion in a dipole-field accretion funnel,
with $\theta_0=0^\circ$.
The system parameters are the same as in Fig.~\ref{fig.rhov}.
}
\label{fig.velocity}
\end{figure}

Figure~\ref{fig.velocity}
  shows the gas velocity
  between the white dwarf surface ($r=1$)
  and the shock.
The velocity profiles show the most significant qualitative difference
  between dipolar and cylindrical accretion models:
  in the Bremsstrahlung-dominated case that we have illustrated,
  the cylindrical model results in a much more constant velocity gradient
  throughout most of the post-shock region.

Figure~\ref{fig.velocity} also displays a significant quantitative difference
  between the predicted shock positions in the cylindrical and dipolar models.
For the cases shown,
  with accretion colatitudes $\theta_0\leq 18^\circ$,
  the shock occurs at a radius around $r_{\rm s}\approx 1.19$.
In the equivalent cylindrical accretion model,
  the shock height is much lower,
  $r_{\rm s}\approx1.105$,
  i.e. just over half the height
  that we would obtain by using the dipole field geometry.

As the magnetic field strength is increased
  to some tens of ${\rm MG}$,
  the radiative cooling due to Cyclotron emission
  becomes large compared to the Bremsstrahlung cooling.
As $\epsilon_{\rm s}$ increases,
  the shock height decreases in both
  the cylindrical accretion model \citep{cropper1999}
  and our dipole model.
For any set of system parameters,
  $(M_*,\dot{m},\theta_0,\epsilon_{\rm s})$,
  the dipolar model predicts the shock to occur higher
  than in the cylindrical model.
However the difference is insignificant 
  in high-$\epsilon_{\rm s}$ cases
  where the shock height 
  is sufficiently low and curvature effects are negligible.
For instances in cases with $B_*=10,30,50{\rm MG}$ 
  we calculate $r_{\rm s}\approx 1.075, 1.016, 1.007$
  according to the cylindrical model,
  but the dipolar model yields
  $r_{\rm s}\approx 1.100, 1.017, 1.007$
  when $\theta_0=0^\circ$.

\begin{figure}
\begin{center} 
\psfig{file=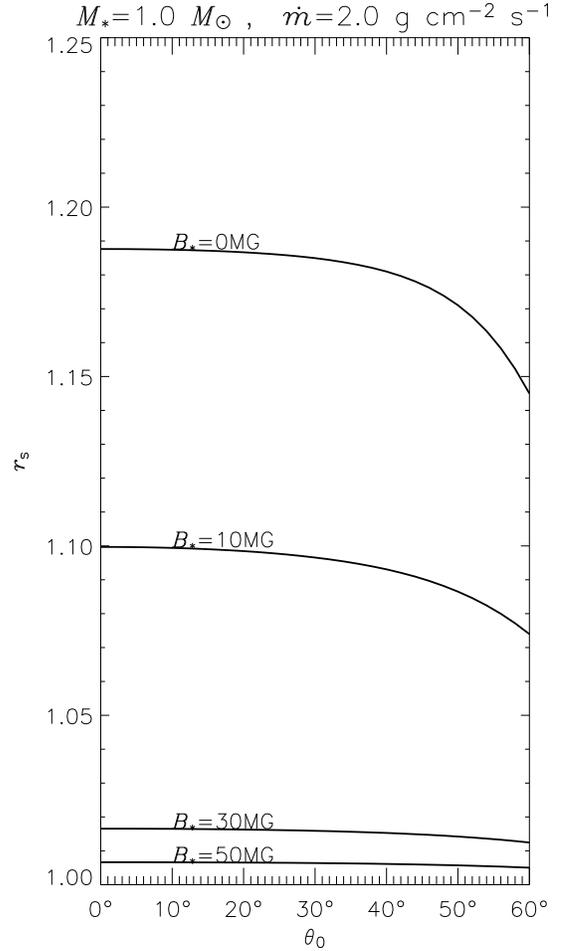,width=8cm}
\end{center}
\caption{
Effect of the accretion colatitude, $\theta_0$,
upon the radial location of the shock,
$r_{\rm s}$ (in units of stellar radius),
for a white dwarf of mass $M_*=1.0M_\odot$
and 
$\dot{m}=2.0\ {\rm g}\,\cm^{-2}\,{\rm s}^{-1}$.
Results are shown for cases with magnetic field strength
$B_* = 0, 10, 30, 50\ {\rm MG}$ at the accretion spot.
}
\label{fig.rshock2}
\end{figure}

Figure~\ref{fig.rshock2} shows the dependence of shock height
  upon the colatitude of the accretion spot ($\theta_0$)
  for several cases of $B_*$
  appropriate for polars.
The maximum shock height generally occurs
  for accretion along the field line with $\theta_0=0^\circ$.
In cases with low $B_*$
  the variation of $r_{\rm s}$ with $\theta_0$
  is less than in the stronger field cases.
When the magnetic field is stronger
  the shock height is smaller compared to the white dwarf radius,
  and the dipolar funnel effects less significant.

\begin{figure}
\begin{center} 
\psfig{file=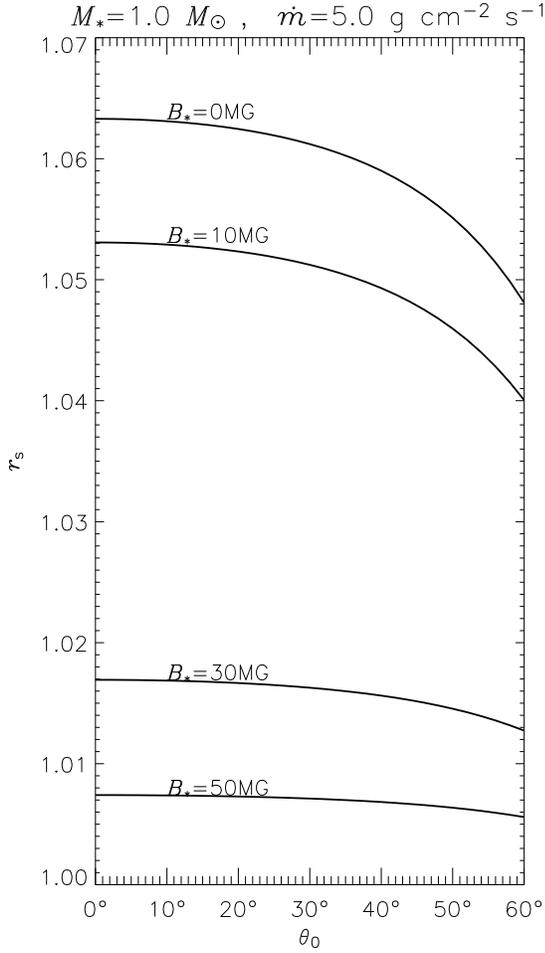,width=8cm}
\end{center}
\caption{
Radial location of the shock as a function of $\theta_0$,
with the same conditions as in Fig.~\ref{fig.rshock2},
except the accretion rate is higher:
$\dot{m}=5.0\ {\rm g}\,\cm^{-2}\,{\rm s}^{-1}$.
}
\label{fig.rshock5}
\end{figure}

Figure~\ref{fig.rshock5} shows equivalent relations between 
  shock height and $\theta_0$,
  but for a greater accretion rate,
  $\dot{m}=5.0\ {\rm g}\,\cm^{-2}\,{\rm s}^{-1}$.
Bremsstrahlung cooling is more efficient in this denser flow,
  and the shock heights are lower than in Fig.~\ref{fig.rshock2},
  as expected from previous studies
  \citep[e.g. ][]{aizu1973,wu1994a,imamura1996}.
Again, as in Fig.~\ref{fig.rshock2},
  $r_{\rm s}$ is more variable in $\theta_0$
  in the cases with weaker magnetic field and greater shock height.

\begin{figure}
\begin{center} 
\psfig{file=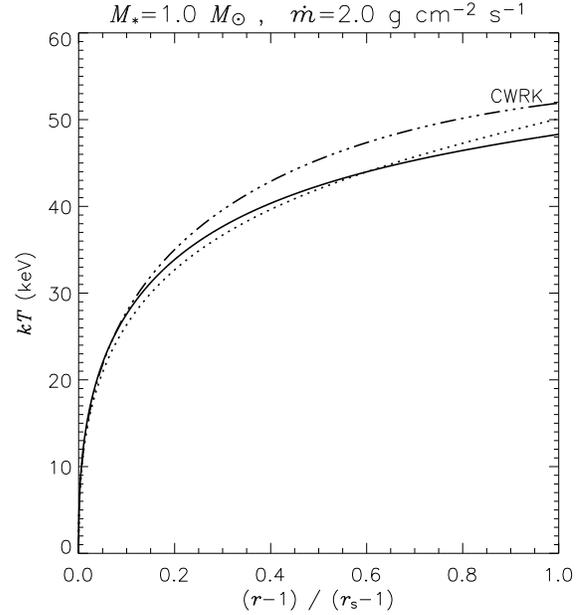,width=8.4cm}
\end{center}
\caption{
Variation of the gas temperature in the accretion column
as a function of altitude scaled to the shock location.
System parameters $(M_*,B_*,\dot{m})$
are the same as in Fig.~\ref{fig.rhov}.
The solid curve is the case of $\theta_0=0^\circ$;
results for $\theta_0\le18^\circ$ appear indistinguishable on this plot.
The nearby dotted curve is the extreme case of $\theta=60^\circ$.
The upper curve (dot-dot-dot-dashed) portrays results from
the cylindrical accretion model (Cropper et al. 1999).
}
\label{fig.T.normal}
\end{figure}

Figure~\ref{fig.T.normal}
  shows the temperature as a function of height within the column
  in our standard illustrative case.
For comparison we calculate the temperature profile
  with the same parameters for the white dwarf and accretion rate,
  but applied in the cylindrical accretion model.
In this specific case
  the shock temperature in the cylindrical accretion model
  is $4\ {\rm keV}$ greater than it is in the dipolar accretion model.
The dipolar accretion model generally yields a lower shock temperature
  than cylindrical accretion,
  because of the greater shock height
  and thus lower free-fall velocity at the shock.
The temperature predictions of cylindrical and dipolar accretion models
  are much more alike for cases where
  the shock is closer to the white dwarf surface
  (and dipolar funnel effects are less significant).

\begin{figure}
\begin{center} 
\psfig{file=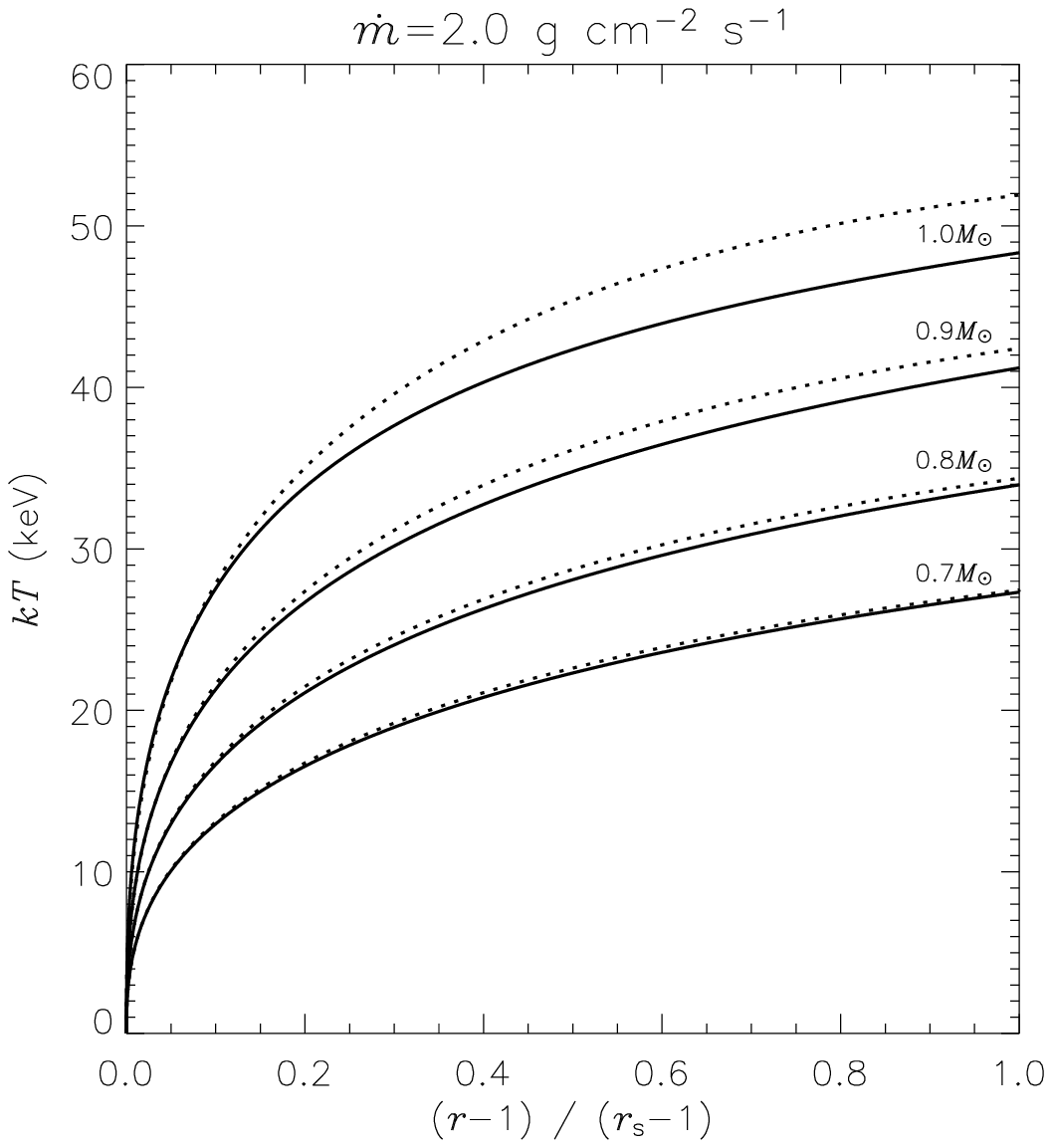,width=8.4cm}
\end{center}
\caption{
Comparison between cylindrical and dipolar geometry
for four different values of white dwarf mass.
From bottom to top each pair of curves represents:
$M_* = 0.7, 0.8, 0.9$ and $1.0M_\odot$.
In each pair the upper (dotted) curve represents
the cylindrical model (CWRK)
and the lower (full line) curve represents the dipolar model.
We take $\theta_0=18^\circ$.
}
\label{fig.T.vs.M}
\end{figure}

Figure~\ref{fig.T.vs.M}
  compares the cylindrical and dipolar models'
  results for the temperature structure of the accretion column,
  for four different values of the white dwarf mass,
  in the Bremsstrahlung-dominated limit.
The dipolar model generally has a lower shock temperature
  than the equivalent cylindrical model.
The difference between models is greatest for the most massive white dwarfs,
  as their accretion shocks are higher
  (and thus more affected by funnel geometry).

In the next section we present calculations of
  the X-ray spectrum emitted from the post-shock flow,
  for two represenative cases of $M_*$.
We compare the predictions of the dipolar accretion model
  with those of the planar models,
  thereby showing the precise effects of curvature.

\subsection{Comments on curvature effects upon the flow}

In section \S~\ref{s.structure}
  we use an accreting magnetic white-dwarf star as an illustration 
  and demonstrate 
  that the role of the field geometry can be as important as 
  (or even more than) the role of gravity  
  in determining the velocity, density and temperature profiles of the flow.   
We can elaborate this more clearly 
  by inspecting the meanings of the terms in the hydrdodynamic equations.   
Each of the terms in the square brackets
  in the right side of equation (\ref{vzetav}) 
  corresponds to an energy-transport process. 
The first represents the radiative loss,  
  the second is determined by the geometry of flow, 
  and the third is due to an external force field, 
  which is only gravity here.  
A characteristic of the second term is the ${\mathcal H}$ function. 
In the curvilinear coordinates that we use, 
  $h_1$ and $h_3$ are the metric elements 
  corresponding to the coordinates perpendicular to $w$.  
Thus ${\mathcal H} (= \partial \ln (h_1 h_3)/\partial w)$ 
  is the measure of the changes in cross-section area 
  of the magnetic-flux tube along the flow, 
  and it plays an important role 
  in determining the efficiency of compressional heating.   

It is worth noting that 
  the ${\mathcal H}$ function also equals to $h_2 \nabla \cdot {\hat b}$, 
  where ${\hat b}$ is the unit vector tangential to the magnetic field,
  and $h_2$ is the metric for the coordinate component along the field line. 
We may also use this relation to determine the ${\mathcal H}$ function.   

For accretion onto stellar objects,  
  $g_{w} \sim \bar{V}_{\rm ff}^2$
  and $\xi \sim \bar{V}_{\rm ff}$, 
  where $\bar{V}_{\rm ff}$ is the mean free-fall velocity.  
The metric element $h_1 \sim {\cal{O}}(1)$, 
  and hence, the geometry function ${\mathcal H} \sim {\cal{O}}(1)$. 
From equation (\ref{vcincomeio}), 
  we can deduce that the effects due to field curvature (geometry) 
  and gravity are often comparable.    
Generally, the geometry effect is more important  
  for regions close to the stellar surface 
  than for regions further away, 
  because the convergence of the dipole field lines.    

We note that equations (\ref{dxidw}) and (\ref {dvdw}) 
  are general equations, 
  in principle, applicable for accretion flows 
  channelled by any field geometries.  
For example, if the magnetic field has planar parallel structure, 
  then we have $h_1 = h_2 = h_3 =1$.      
It follows that ${\mathcal H}=0$.   
In this flow, the accreting material will not be compressed by the magnetic field.    
If the flow is spherical,  
  then $h_1=1$, $h_2 = r$, $h_3 = r \sin \theta$ (see e.g.\ \citealt{arfken}),
  and ${\mathcal H} = 2/r$.   
The flows channelled by dipole field have been shown in the sections above. 
The generalisation to channelled flows in higher-order fields
  is possible.
What one needs are first 
  to find the functional expressions of the field lines and the equipotential surfaces, 
  and use them to define the coordinates. 
After the coordinate system is specified, 
  we can derive the metric elements $h_1$, $h_2$ and $h_3$
  and determine the ${\mathcal H}$ function.    

The effects of more complex field geometries depend on
  the alignment and distribution of the higher-order field components.
If the field lines near the accretion spot diverge more rapidly
  than in the dipolar model (i.e. a more flared accretion funnel)
  then we expect an accentuation of the 
  effects found in the present work:
  a higher shock position and lower shock temperature.
If the field lines near the accretion spot diverge less rapidly
  then the accretion flow will be a closer approximation to
  the older, planar accretion models.


\section{X-ray emission}

\subsection{Spectral calculations}

We use the analytic model 
to provide electron density and temperature values
throughout the post-shock flow.
The structure is divided into between $2\times10^3$ and $2\times10^4$ strata,
depending on the adaptive integration steps
chosen by the routine that solves the set of differential equations.
Using these $(n_{\rm e}, T_{\rm e})$ values,
an XSPEC subroutine
for a MEKAL model of an optically thin plasma
\citep{mewe1995,phillips1999}
calculates the X-ray emission spectrum of each layer.
The total X-ray spectrum of the accretion column
is a sum over all the layers' spectra,
weighted by cell volume.
We omit the few cells for which $n_{\rm e}>1.0\times10^{18}\,\cm^{-3}$,
for which the MEKAL model is inapplicable and fails.
In reality the gas at these densities is optically thick 
and merges into the stellar atmosphere.

\begin{figure}
\begin{center} 
\psfig{file=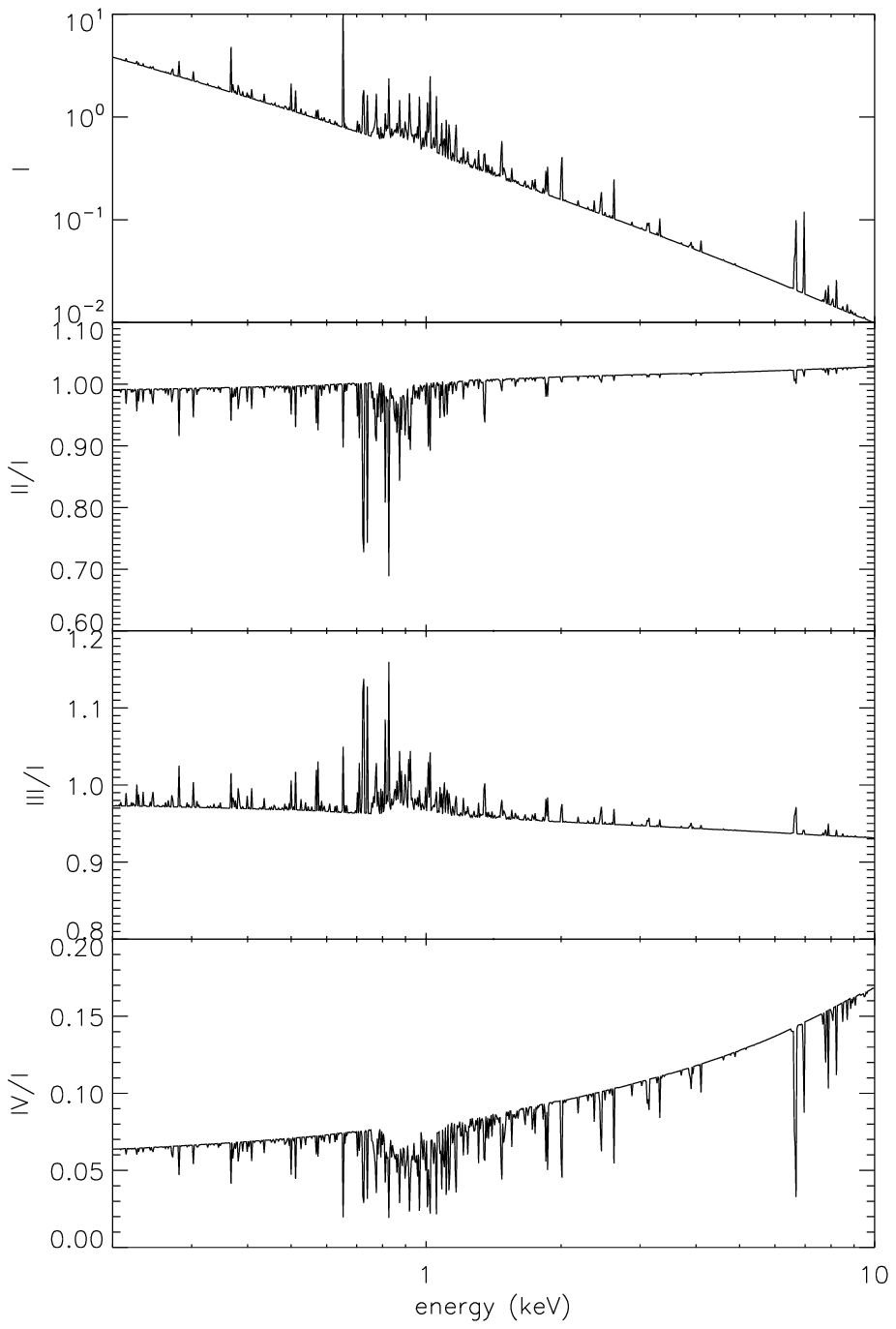,width=8.4cm}
\end{center}
\caption{
Comparison of X-ray spectra calculated using
(I) the dipole accretion model with gravity;
(II) CWRK accretion model with gravity but without curvature effects;
(III) planar accretion model of Wu (1994) without gravity or curvature;
(IV) isothermal model using the post-shock temperature and density
derived from model (I).
The upper panel shows the X-ray spectrum of model (I) directly.
The remaining panels present ratios between
the dipole model spectrum and the other models.
In all cases we set
$\dot{m}=2.0\ {\rm{g}}\ \cm^{-2}\ {\rm{s}}^{-1},
 B_*=10\ {\rm MG}, 
 M_*=1.0M_\odot,
 \theta_0=0^\circ$.
}
\label{fig.spectra.1.0}
\end{figure}

Figure~\ref{fig.spectra.1.0}
presents the X-ray spectrum 
calculated for a high-mass white dwarf ($1.0\Msun$)
with a magnetic field strength of $10\ {\rm{MG}}$
at the accretion hot-spot.
The uppermost panel represents
the dipolar accretion model (I).
The lower panels are ratio plots, 
comparing the dipolar model with:
(II) the model of \cite{cropper1999}
(planar geometry, including gravity);
(III) the model of \cite{wu1994a} 
(planar, but without gravity);
and (IV) an isothermal, homogeneous post-shock model.
For photon energies ranging from $0.2$ to $10~{\rm{keV}}$,
the spectrum of model (II) is harder than that of the dipolar model (I),
and has less line emission.
It may be unsurprising that the CWRK model has a harder spectrum,
since the shock temperature is higher than in the dipolar accretion model.
The planar accretion model without gravity (III)
produces a softer spectrum than the dipolar accretion model,
and more line emission.
The homogeneous shock model (IV) yields a spectrum that is
qualitatively very different from any of the inhomogeneous shock models.
It shows a much harder spectal slope.

\begin{figure}
\begin{center} 
\psfig{file=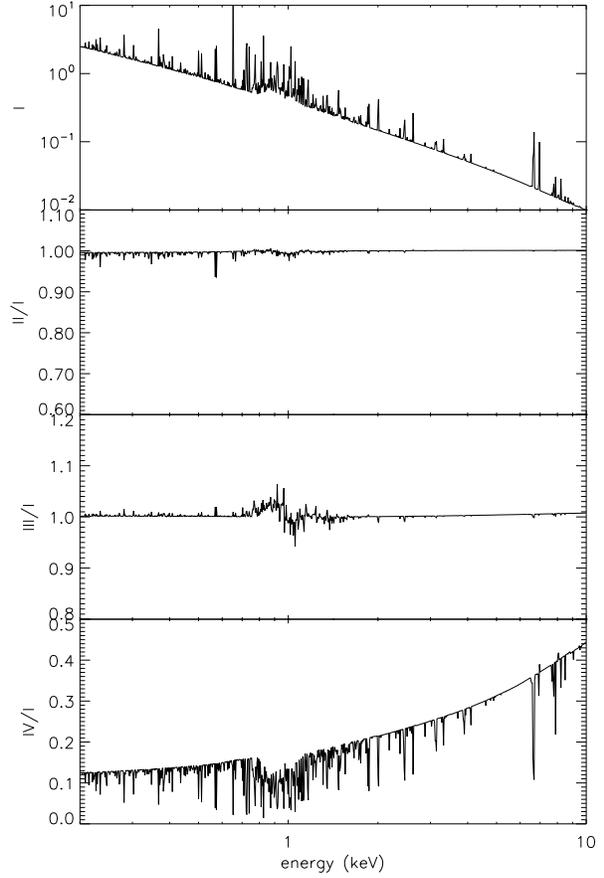,width=8.4cm}
\end{center}
\caption{
Comparison of X-ray spectra calculated according to the four models,
labelled as in Fig.~\ref{fig.spectra.1.0}
but with stellar mass
$M_*=0.5M_\odot$.
}
\label{fig.spectra.0.5}
\end{figure}

Curvature effects are strongest for those cases
where the shock height is significant compared to the stellar radius,
and thus the differences between spectra in comparable 
CWRK and dipolar accretion models
is most noticeable for the cases of the most massive white dwarfs.
For lower mass white dwarfs the shock height is small
compared to the stellar radius,
curvature effects are minor,
and the dipolar and CWRK models predict similar X-ray spectra.
For the case of 
$M_*=0.5M_\odot$
(Fig.~\ref{fig.spectra.0.5})
the spectra predicted from these models are almost indistinguishable.


\subsection{Effects on X-ray spectrum}

For white dwarfs of low or medium mass
(e.g. $0.5M_\odot$)
the shock temperatures, post-shock structures and X-ray spectra
are nearly the same in the dipolar accretion model
as in the earlier planar accretion model of \cite{cropper1999}.
For white dwarfs of higher mass
(e.g. $1.0M_\odot$),
the dipolar accretion model predicts
greater shock heights,
lower shock temperatures
and softer X-ray spectra 
in the band from $0.2 - 1.0~{\rm{keV}}$.
This implies that for a given observed spectrum,
the mass estimate $M_*$ is necessarily greater
when dipole field curvature is taken into account.

The lower shock temperature does not at first sight provide
  a route to resolving the high white dwarf masses
  derived using spectral fits to X-ray data,
  and which are the subject of some controversy
  \citep[e.g.][]{schwope2002},
  since even higher mass white dwarfs are required
  to provide sufficient hard X-ray flux to fit the observed spectrum.

\section{Application to Other Systems}

The formulation that we derived is not restricted
   to accretion in magnetic white dwarfs.
It is applicable to a variety of astrophysical systems,
   provided that the flow is strictly confined by the magnetic field lines,
   and the radiative heating and cooling processes can be
   parametrised in terms of the hydrodynamic variables.

An example is accretion in young stellar objects.
There are strong observational evidence
  that accretion flow in T~Tauri stars are channelled by
  the stellar magnetic fields (e.g.\ \citealt{basri1992}).
The main differences between the mCV case
  and the accretion in T~Tau stars are:
  that the cooling process is dominated by
  line cooling instead of free-free and cyclotron cooling.
Moreover,  irradiative heating of both pre- and post-shock flow
  is important in determining the flow hydrodynamics
  (see e.g.\ \citealt{martin1996}).
Under these conditions,
  we need to relax the assumption of a completely dissociated gas
  and generalise the hydrodynamical equations to multi-ion and electron flow.
In a fully self-consistent treatment
  the equations of hydrodynamics
  need to be solved simultaneously with
  the ionization structure equations
  \citep[e.g. ][]{lamzin1998}.

Another example of application is
  the accretion onto a slowly rotating neutron star in a binary system
  (i.e. X-ray pulsars).
The accretion flow is magnetically funnelled,
  from either the wind of the companion star
  or an accretion disk
  \citep[e.g.][]{elsner1977,ghosh1978,lovelace1995,koldoba2002}.
In cases of disk accretion,
  the gas is funnelled from the inner edge of the disk,
  and the radius of this location effectively determines
  the colatitude of the accretion hot-spot on the stellar surface
  ($\theta_0$ in our formulation).
For details of interaction of the channelled flow
  and the accretion disk in X-ray pulsars, see e.g.
  \citet{anzer1983,spruit1990,li1996b}.

\section{Summary}  
 
We have investigated accretion onto stellar systems   
   in which the flow is strictly channelled by a dipole magnetic field.     
Such flow occurs in regions close to the stellar surface,  
   where $B \gg 8\pi \rho v^2$. 
We derive a set of hydrodynamic equations 
   using the curvilinear coordinates natural to the field geometry. 
The equations are solved 
   to determine the flow velocity, density and temperature profiles.   
We show that the dipole-field geometry 
   can cause significant compressional heating, 
   and this effect can be comparable to radiative cooling and gravity 
   in determining the structures of the flow 
   near the surface of the accreting star.   

The formulation that we have derived
   is applicable to a variety of astrophysical systems, 
   from white-dwarf stars in magnetic cataclysmic variables  
   to young stellar objects.   
It would be possible
   to generalise the formulation for dipole-field channelled flow 
   to the flow channelled by a higher-order field 
   using curvilinear coordinates.  

We demonstrate that our analytic model can be 
  efficiently interfaced with spectral software 
  to reproduce fits in analysis of high quality X-ray spectra,
  obtained by satellites such as {\em Chandra} and {\em XMM-Newton}.
This is a next-generation model
  for fitting accretion parameters, 
  including white dwarf mass, magnetic field strength,
  accretion colatitude and mass flux,
  to specific observed systems.

\begin{acknowledgements}
JBGC is very grateful 
  for the great hospitality of all people 
  of the Mullard Space Science Laboratory (MSSL) 
  and particularly to Gavin Ramsay and Mark Cropper 
  for the invitation to visit MSSL for one year. 
JBGC also thanks Roberto~Soria, Mat~Page and Christian~Bridge 
  for discussions.
JBGC acknowledges the support 
  from the Conselho Nacional 
  de Desenvolvimento Cient\'{\i}fico e Tecnol\'{o}gico (CNPq) of Brazil 
  and from the State University of Rio de Janeiro (UERJ) 
  for the one-year leave inside the PROCAD program .
\end{acknowledgements}



\appendix

\section{Orthogonal Basis and Metric Elements}

Let $\BHi$, $\BHj$ and $\BHk$
  be the standard orthogonal unit vectors in the Cartesian coordinates.  
The unit vectors of our dipolar curvilinear coordinate system
  ($\BHu, \BHw, \BHphi$)  
  and ($\BHi$, $\BHj$, $\BHk$) 
  are related to the Cartesian unit vectors via
\begin{equation}
\left[\begin{array}{c}
  \hat{\mathbf u}\\
  \hat{\mathbf w}\\
  \BHphi
  \end{array}\right]
   = {\mathsf U} \left[\begin{array}{c}
  \BHi\\
  \BHj\\
  \BHk
   \end{array}\right]  \ .   
\end{equation}  
where (see e.g.\ \citealt{arfken})
  the transformation matrix is
\begin{equation}
  {\mathsf U}\!=\!
\left[\begin{array}{ccc}
\frac{\displaystyle 1}{\displaystyle h_{1}}\frac{\displaystyle \partial x}{\displaystyle \partial u}     &
\frac{\displaystyle 1}{\displaystyle h_{1}}\frac{\displaystyle \partial y}{\displaystyle \partial u}     &
\frac{\displaystyle 1}{\displaystyle h_{1}}\frac{\displaystyle \partial z}{\displaystyle \partial u}\\
\frac{\displaystyle 1}{\displaystyle h_{2}}\frac{\displaystyle \partial x}{\displaystyle \partial w}     &
\frac{\displaystyle 1}{\displaystyle h_{2}}\frac{\displaystyle \partial y}{\displaystyle \partial w}     &
\frac{\displaystyle 1}{\displaystyle h_{2}}\frac{\displaystyle \partial z}{\displaystyle \partial w}\\ 
\frac{\displaystyle 1}{\displaystyle h_{3}}\frac{\displaystyle \partial x}{\displaystyle \partial \varphi} &
\frac{\displaystyle 1}{\displaystyle h_{3}}\frac{\displaystyle \partial y}{\displaystyle \partial \varphi} &
\frac{\displaystyle 1}{\displaystyle h_{3}}\frac{\displaystyle \partial z}{\displaystyle \partial \varphi} 
  \end{array}\right] \ . 
\label{matrix U definition}
\end{equation}    
The coefficients $h_{1}$, $h_{2}$ and $h_{3}$ 
  are the metric of the curvilinear coordinates system, 
  and they are 
\begin{equation}
  h_{[1,2,3]} = \sqrt{   \left(\frac{\partial x}{\partial [u,w,\varphi]}\right)^{2}
         +\left(\frac{\partial y}{\partial [u,w,\varphi]} \right)^{2}
         +\left(\frac{\partial z}{\partial [u,w,\varphi]}\right)^{2} }   \ , 
\label{h1 definition}
\label{h2 definition}
\label{h3 definition}
\end{equation}

The transformation from between dipolar coordinates
  and spherical coordinates 
  ($\BHr, \BHtheta, \BHphi$)  
  is given by:
\begin{equation}
\left[\begin{array}{c}
  \BHr\\
  \BHtheta\\
  \BHphi
  \end{array}\right]
  = {\mathsf N} \left[\begin{array}{c}
  \BHi\\
  \BHj\\
  \BHk
  \end{array}  \right]
= {\mathsf Q} \left[\begin{array}{c}
    \BHu\\
    \BHw\\
    \BHphi
  \end{array}  \right] \ ,
\end{equation}
  where the preceding matrix is
\begin{equation}
  {\mathsf N}=\left[\begin{array}{ccc}
  \sin\,\theta \,\,\cos\,\varphi& \sin\,\theta \,\,\sin\,\varphi &  \cos\,\theta\\
  \cos\,\theta \,\,\cos\,\varphi& \cos\,\theta \,\,\sin\,\varphi &  -\sin\,\theta\\
  -\sin\,\varphi&\cos\,\varphi &  0
   \end{array}\right] \ , 
\label{matrix N definition}
\end{equation}
  ${\mathsf Q} = {\mathsf N} \cdot {\mathsf M}$,
  and
  ${\mathsf M}\equiv {\mathsf U}^{t}={\mathsf U}^{-1}$
  (the transpose of ${\mathsf U}$),
Thus ${\mathsf Q}$ completely prescribes the relationship between
  ($\BHu, \BHw, \BHphi$)   
  and ($\BHr, \BHtheta, \BHphi$).

  
With $u$ and $w$ defined according to (\ref{udipole}) and (\ref{wdipole})
  we have $\sin^2\theta = ur$ and $\cos^2\theta =w^2 r^4$,
  and as $\cos^{2}\,\theta + \sin^{2}\,\theta = 1$, 
  we therefore have 
\begin{equation}
   w^{2}r^{4} + u\,r -1 = 0 \ .  
   \label{quartic.w.u}
\end{equation} 
This is a quartic equation, 
  which has two real and symmetrical roots 
  and two imaginary roots 
  (e.g.\ \citealt{abramowitz}).
The real root that is of physical relevance is 
\begin{equation}
    r=\sqrt{Y-S_+-S_-}-{u\over{4w^2Y}}
    \label{eq.radius.fn}
    \label{ruv}
\end{equation}
   where 
\begin{equation}
Y=\sqrt{(S_++S_-)^2+3(S_+-S_-)^2}
\end{equation}
\begin{equation}
S_\pm \equiv\left({ -\half{W}\pm\sqrt{X} }\right)^{1/3}
\ ,
\end{equation}
\begin{equation}
W=-{{u^2}\over{64w^4}}
\ \ \ \ \ \mbox{and}
\end{equation}
\begin{equation}
X
=\left({1\over{128w^4}}\right)^2
\left({
u^4 + {{256w^2}\over{27}}
}\right)
\ .
\end{equation}

The first derivatives of the radius function,
$r'_{u}\equiv\partial r/\partial u$
and $r'_{w}\equiv\partial r/\partial w$,
are obtainable by implicit differentiation of (\ref{quartic.w.u}):
\begin{equation}
r'_{u}=
{{-r}\over{4w^2r^3+u}}
\ \mbox{\ \ and}
\label{drdu}
\end{equation}
\begin{equation}
r'_{w}=
{{-2wr^4}\over{4w^2r^3+u}}
\ .
\label{drdw}
\end{equation}

The relation between ($x$, $y$, $z$) and ($u$, $w$, $\varphi$) is given by
equations (\ref{assign.xy}) and (\ref{assign.z}).
It follows that  
\begin{equation}
  {\partial\over{\partial u}}
  \left[{\begin{array}{ccc}
  x\ & y 
  \end{array}}\right]
  =
     \frac{1}{2} \sqrt{\frac{r}{u}}\, (r + 3ur'_{u})\,
  \left[{\begin{array}{ccc}
     \cos\,\varphi\ & \sin\,\varphi
  \end{array}}\right]\ ,
\label{xu}
\label{yu}
\end{equation}
\begin{equation}
  \frac{\partial z}{\partial u} 
    = 3\, w\, r^{2} r'_{u} \mbox{,}
\label{zu}
\end{equation} 
\begin{equation}
  {\partial\over{\partial w}}
  \left[{\begin{array}{ccc}
  x\ & y 
  \end{array}}\right]
  =
     \frac{1}{2} \sqrt{\frac{r}{u}}\, (r + 3ur'_{u})\,
  \left[{\begin{array}{ccc}
     \cos\,\varphi\ & \sin\,\varphi
  \end{array}}\right]\ ,
\label{xw}
\label{yw}
\end{equation}
\begin{equation}
  \frac{\partial z}{\partial w} 
    = \, r^{2}(r + 3\,w\,r'_{w})  \mbox{,}
\label{zw}
\end{equation}
\begin{equation}
  {\partial\over{\partial\varphi}}
  \left[{\begin{array}{ccc}
  x\ &y\ &z
  \end{array}}\right]
  =
  \sqrt{ur^3}
  \left[{\begin{array}{ccc}
  -\sin\varphi\ &\cos\varphi\ &0
  \end{array}}\right]
\ .
\label{xfi}
\label{yfi}
\label{zfi}
\end{equation} 

Consider these substitions into (\ref{h1 definition}):
  (i) equations (\ref{xu}) and (\ref{zu}) 
  into $h_1$;
  (ii) equation (\ref{xw}) and (\ref{zw}) 
  into $h_2$;
  and 
  (iii) equation (\ref{xfi}) 
  for $h_3$;
Then we obtain metric elements explicitly:
\begin{equation}
   h_{1} = \sqrt{
    \frac{r}{4\,u}(r + 3\, u\, r'_{u})^{2}
      +(3\,w\,r^{2}\,r'_{u})^{2}} \mbox{,}
\label{h1}
\end{equation}
\begin{equation}
   h_{2} = \sqrt{r^{4}\,(r+3\,w\,r'_{w})^{2}
     +{{\textstyle\frac{9}{4}}} u\,r r_{w}^{'2}
    } \mbox{,}
\label{h2}\end{equation}
\begin{equation}
   h_{3} = \sqrt{u\,r^{3}} \mbox{.}
\label{h3}
\end{equation}

It is also worth noting that 
  $\BHu, \BHw$ and $\BHphi$
  vary along the corresponding coordinate axes. 
It can be shown that  
\begin{equation}
  \frac{\partial \hat{\mathbf u}}{\partial u}
     = -\left(\frac{\hat{\mathbf w}}{h_{2}}\,
      \frac{\partial h_{1}}{\partial w} + \frac{\BHphi}{h_{3}}\,
      \frac{\partial h_{1}}{\partial \varphi} \right) \mbox{,}
\label{delh1fi}
\end{equation}
\begin{equation}
  \frac{\partial \hat{\mathbf w}}{\partial w} 
     = -\left(\frac{\hat{\mathbf u}}{h_{1}}\,
      \frac{\partial h_{2}}{\partial u} + \frac{\BHphi}{h_{3}}\,
      \frac{\partial h_{2}}{\partial \varphi} \right) \mbox{,}
\label{delh2fi}
\end{equation}
\begin{equation}
  \frac{\partial \BHphi}{\partial \varphi}
     = -\left(\frac{\hat{\mathbf u}}{h_{1}}\,
      \frac{\partial h_{3}}{\partial u} + \frac{\hat{\mathbf w}}{h_{2}}\,
      \frac{\partial h_{3}}{\partial w} \right)
\label{delh3w}
\end{equation}
 (see e.g.\ \citealt{arfken}).

\section{The ${\mathcal H}$ Function} 

The function ${\mathcal H}(u,w)$ is a purely geometrical function, 
  dependent on the curvature of the coordinate system. 
In the ($u$, $w$, $\varphi$) coordinate systems that we use,  
  it is 
\begin{equation}
    {\mathcal H}= \frac{3\,u\,r^{5}}{(h_{1}h_{3})^2} \left[ 
       \frac{r'_{w}\,u}{4}\left( \frac{1}{u} 
      + \frac{3\, r'_{u}}{r}\right)t_{7}
      + 3\, \left(  w\, r\, r'_{u}\right)^{2} t_{8}\right] \mbox{,}
\label{hexplicity}
\end{equation}
   where
\begin{equation}
  t_{7} = \frac{1}{u} + \frac{2\, r'_{u}}{r} 
     + \frac{r''_{uw}}{r'_{w}} \mbox{,}
\label{t7}\end{equation}
\begin{equation}
  t_{8 }= \frac{1}{w} + \frac{7}{2}\frac{r'_{w}}{r} 
      + \frac{r''_{uw}}{r'_{u}} \mbox{.}
\label{t8}
\end{equation}
In the above expression, 
   $r''_{uw}=\partial r'_{u}/\partial w 
   = \partial(\partial r/\partial u)/\partial w$, 
   which is obtained by implicit differentiation of (\ref{drdu}):
\begin{eqnarray}
r''_{uw}&=&
-{{\left[{
r'_w + \left({ 8wr^3+12w^2r^2 r'_w }\right) r'_u
}\right]
}\over{ 4w^2r^3 + u }}
\ .
\label{ddrdwdu}
\end{eqnarray}

\begin{figure}
\begin{center} 
\vspace*{2.5cm}  
\psfig{file=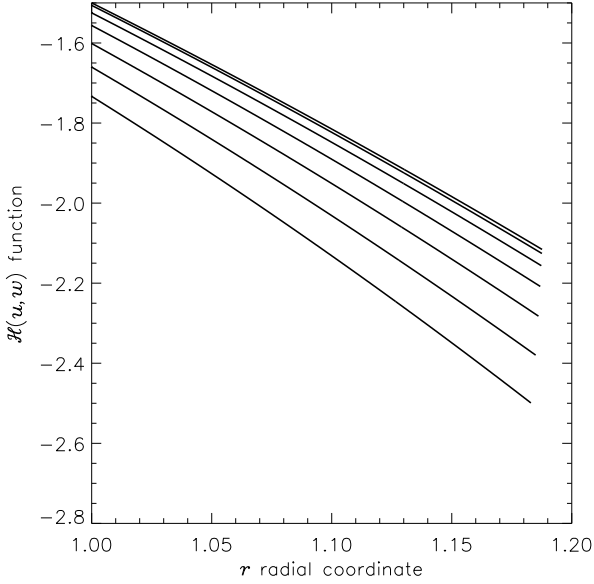,width=8.4cm}
\end{center}
\caption{
From top to bottom, the ${\mathcal H}(u,w)$ function ( 
   for $\theta_o = 0^\circ, 6^\circ, 12^\circ, 18^\circ, 24^\circ, 30^\circ$
   and $36^\circ$ respectively.  
The horizontal axis is the radial distance from the centre of the star.
In the calculation 
   $\dot{m} = 2.0\, {\rm g}\,\cm^{-2}\,{\rm s}^{-1}$ 
   and $M_*=1.0\,{\rm M}_{\odot}$.  }
\label{plot of h function}
\end{figure}

We can also relate the change in $\rho\,v$ to the ${\mathcal H}(u,w)$ function. 
The mass-continuity equation implies 
\begin{equation}
   h_{1}h_{3}\frac{\partial}{\partial{w}}(\rho\,v)+
   \rho\,v\frac{\partial}{ \partial{w} }( h_{1}h_{3} ) = 0 \ . 
\end{equation} 
Hence, 
\begin{equation}
   \frac{\partial}{\partial{w}}ln(\rho\,v) = -{\mathcal H} \ . 
\end{equation}
Figure (\ref{plot of h function}) 
  shows the ${\mathcal H}(u,w)$ function 
  for some values of $\theta_o$, 
  the angle between the axis of the dipole 
  and the position vector of the foot point of the magnetic-field line. 

\section{Cyclotron cooling efficiency $\epsilon_{\rm s}$}
\label{app.epsilons}

In our illustrative system
the Cyclotron emission from the post-shock region
is optically thick at frequencies up to some cut-off,
$\omega_*$
\citep[e.g. ][]{chanmugam1979,wada1980,langer1982}.
As derived in \citet{saxton.thesis}
and the appendices of \citet{cropper1999},
the effective volumetric cooling function is
\begin{equation}
\Lambda_{\rm cy}
={{\pi D}\over{a}} {{k_{\rm B}T}\over{12\pi^2c^2}} \omega_*^3
\ ,
\end{equation}
  where
  $\omega_*\approx9.87\omega_{\rm c}
  (\Theta/10^7)^{0.05}(T/10^8{\rm K})^{0.5}$,
  $\omega_{\rm c}=eB/m_{\rm e}c$ is the Cyclotron frequency,
  $\Theta=2\pi e n_{\rm e} D / B$,
  $D$ is the diameter of the accretion flow,
  $a=\pi D^2/4$ is the cross-sectional area,
  $B$ is the magnetic field strength,
  $T$ and $n_{\rm e}$
  are the electron number density and temperature respectively.
Calculating $\Lambda_{\rm cy}$
  in the conditions immediately downstream of the shock,
  and taking its ratio with the Bremsstrahlung cooling
  (\ref{Lambda.ff})
  gives the relative efficiency parameter
  $\epsilon_{\rm s}\equiv \Lambda_{\rm cy,s}/\Lambda_{\rm br,s}$
  \citep{wu1994a}.
An appropriate series of substitutions leads to (\ref{def.f}).

However there is a practical problem to performing calculations
  starting from a chosen value of $\epsilon_{\rm s}$,
  since (\ref{def.f})
  depends on functions of the shock position,
  which is initially unknown.
In the earliest cylindrical accretion models
  \citep{aizu1973,chevalier1982,wu1994a}
  it was possible to eliminate the shock height by normalisation,
  but this is impossible for formulations that include
  explicitly position-dependent effects such as gravity 
  \citep{cropper1999}.
For computational convenience, we modify the definition by
  replacing the ${\rm s}$-subscripted quantities in (\ref{def.f}) with
  reference values at the stellar surface
  ($B_*$, $h_{1*}$, $h_{3*}$),
  and ideal post-shock pressure and densities
  $(P'_{\rm s},\rho'_{\rm s})$
  of the zero-gravity, cylindrical model of \citet{wu1994a}.
In that model the shock was assumed to occur close to the stellar surface.
The corresponding efficiency parameter is denoted
  $\epsilon_{{\rm s}*}$,
  and it yields, after solving for the actual shock location,
  the value of 
\begin{equation}
\epsilon_{\rm s}
=
\epsilon_{{\rm s}*}
\left({
{P_{\rm s}}\over{P'_{\rm s}}
}\right)^2
 \left({
	{\rho'_{\rm s}}\over{\rho_{\rm s}}
 }\right)^{\frac{77}{20}}
 \left({
	{{h_{1*}h_{3*}}\over{h_{1\rm s}h_{3\rm s}}}
 }\right)^{\frac{17}{40}}
 \left[{ 
	{{4-3ur_{\rm s}}\over{(4-3u)r_{\rm s}^6}}
 }\right]^{\frac{57}{40}}.
\end{equation}

In our main illustrative case,
  with
  $M_*=1.0M_\odot$,
  $V_*=6.910\times10^{8}\,\cm\,{\rm s}^{-1}$,
  $\dot{m}=2.0\ {\rm g}\,\cm^{-2}\,{\rm s}^{-1}$
  and approximately solar abundances,
  we derive
  $\epsilon_{\rm s*}=3.2, 72.7, 312$
  and
  $\epsilon_{\rm s }=1.6, 64.6, 297$
  for $B_* = 10, 30, 50\ {\rm MG}$ respectively.

\end{document}